\documentclass[
twocolumn,
prl,showpacs,preprintnumbers,amsmath,amssymb]{revtex4}

\usepackage{graphicx}
\usepackage{amssymb}
\usepackage{amsmath}
\usepackage{dsfont}
\usepackage{bm}
\usepackage{mathrsfs}

\begin{document}


\title{An exact solution for the Casimir force in a spherically symmetric medium}
\author{Ulf Leonhardt and William M.\ R.\ Simpson}
\affiliation{
School of Physics and Astronomy, University of St Andrews, 
North Haugh, St Andrews KY16 9SS, UK
}
\date{\today}
\begin{abstract}
We calculated the force of the quantum vacuum, the Casimir force, in a spherically symmetric medium, Maxwell's fish eye, surrounded by a perfect mirror and derived an exact analytic solution. Our solution questions the idea that the Casimir force of a spherical mirror is repulsive --- we found an attractive force that diverges at the mirror. 
\end{abstract}
\pacs{03.70.+k, 77.84.Lf}
\maketitle

Casimir suggested an intriguing model that could explain the stability of charged particles and the value of the finestructure constant \cite{Casimir1}. The argument goes as follows: Imagine the particle as an electrically charged hollow sphere. Two forces are acting upon it: the electrostatic repulsion and the force of the quantum vacuum, the Casimir force --- presumed to be attractive \cite{Casimir0,Lifshitz,Milonni,Milton,BKMM,Review,Leo}. The stress $\sigma$ of the quantum vacuum on a spherical shell of radius $a$ must be given by a dimensionless constant times $\hbar c/a^4$ on purely dimensional grounds --- the quantum stress is an energy density proportional to $\hbar$, and $\hbar c/a^4$ carries indeed the units of an energy density. Now, the electrostatic energy of the sphere is proportional to the square $e^2$ of its charge and is also inversely proportional to $a^4$ \cite{Jackson}. Therefore, an attractive Casimir force balances the electrostatic repulsion regardless of how small $a$ is, provided $e^2/(\hbar c)$ assumes a certain value given by the strength of the Casimir force. This strength depends on the internal structure of the particle --- the fact that it is a spherical shell --- but not on its size, which could be imperceptibly small. Casimir's model, however crude, could simultaneously explain the finestructure constant $e^2/(\hbar c)$ and the stability of charged elementary particles!

All one needs to do is calculate the Casimir force on a spherical shell, but such calculations are notoriously difficult. After a marathon struggle with special functions, Boyer succeeded in numerically computing the force for an infinitely conducting, infinitely thin shell and found a surprising result \cite{Boyer} that shattered Casimir's idea: the vacuum force is repulsive and so cannot possibly balance the electrostatic repulsion. Boyer's heroic calculation was confirmed in a sophisticated and elegant paper by Milton, DeRaad and Schwinger \cite{MDS} and by others \cite{Other}. The spherical shell has become the archetype for a shape that causes Casimir repulsion \cite{Levitationremark}. However, doubts have been lingering about whether the repulsive force of the shell may be an artefact of the simple model used \cite{Doubts}, for the following reason: the bare stress of the quantum vacuum is always infinite and this infinity is removed by regularization procedures \cite{Casimir0,Lifshitz,Milonni,Milton,BKMM,Review,Leo}. The most plausible regularization involves considering the relative stress between or inside macroscopic bodies. But an infinitely thin sphere does not represent an extended macroscopic body, nor multiple bodies. Suppose the physically relevant vacuum stress of an extended spherical shell tends to infinity in the limit when the shell becomes infinitely thin and infinitely conducting. In this case the regularization would remove this physically significant infinity, producing a finite result that may very well have the wrong sign. Our paper supports the contention that the Casimir repulsion of the spherical shell could be an artefact of regularization. 

\begin{figure}[t]
\begin{center}
\includegraphics[width=17.0pc]{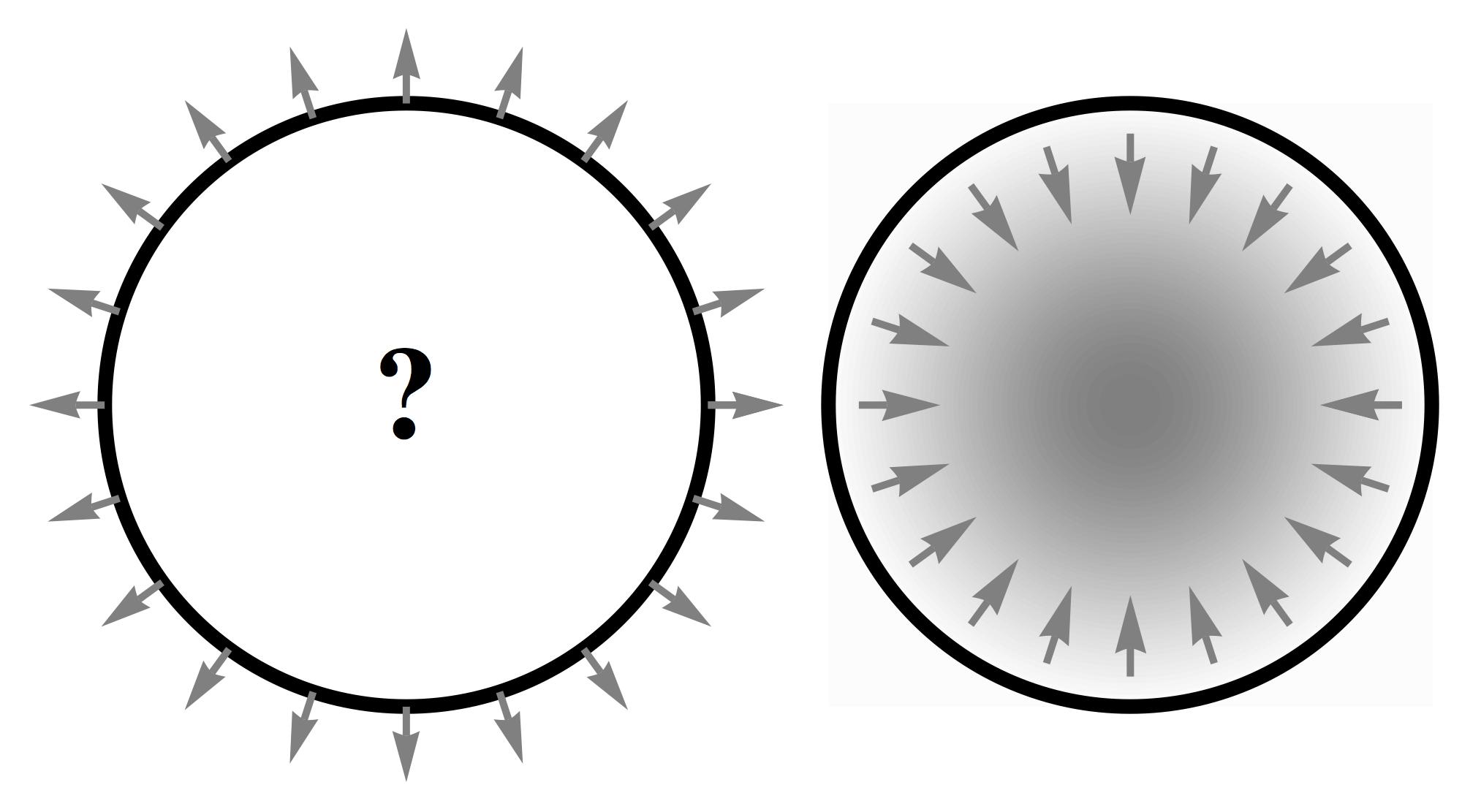}
\caption{
\small{
The Casimir force on a spherical shell (left) is repulsive \cite{Boyer}, or is it? We assumed the shell to be filled with a medium (right) and found an attractive force. The shades of grey indicate the profile of the medium (plotted in Fig.~2).}
\vspace*{-5mm}
}
\end{center}
\end{figure}

\begin{figure}[h]
\begin{center}
\includegraphics[width=17.0pc]{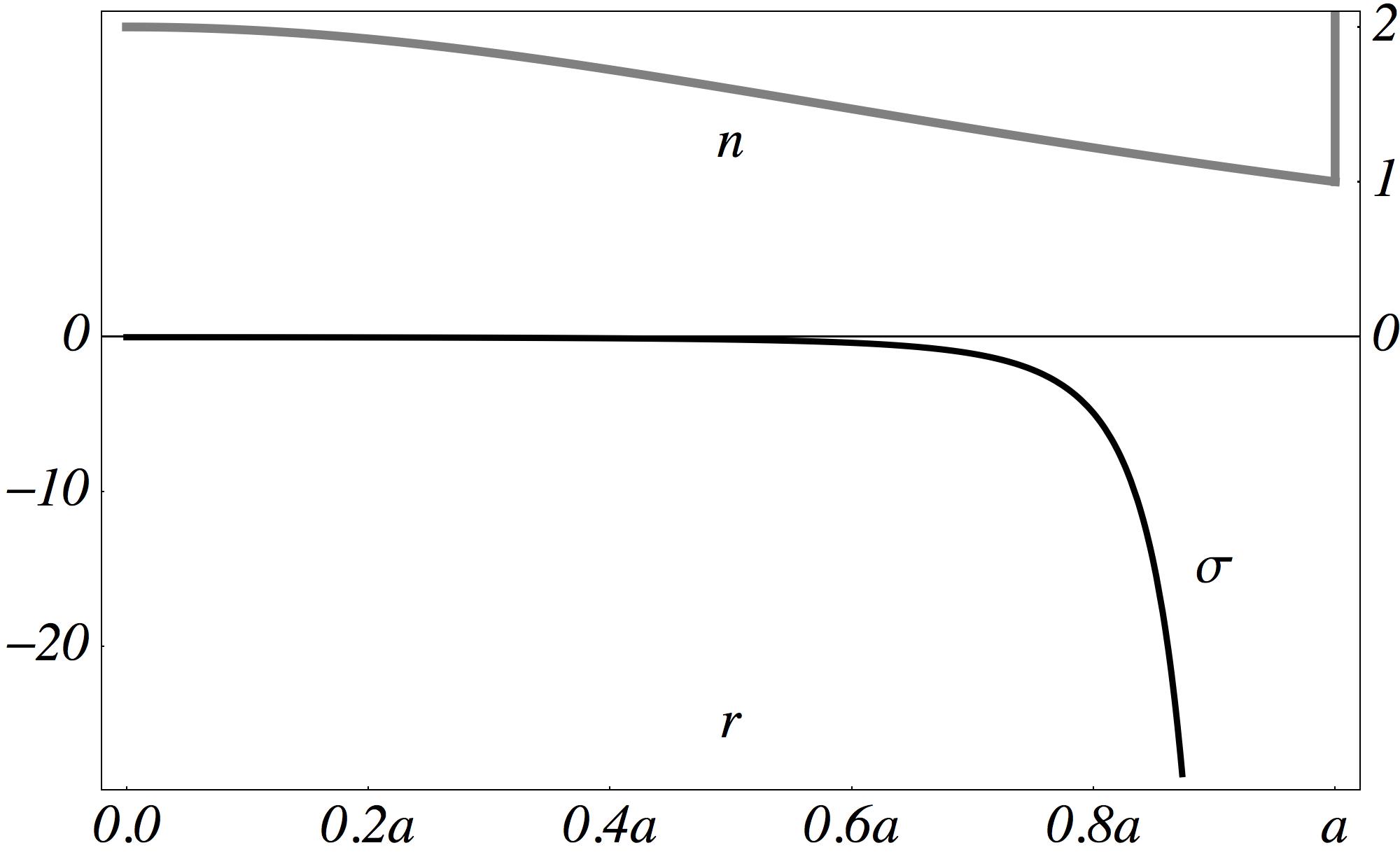}
\caption{
\small{
Index profile $n(r)$ (grey curve) of the medium inside the shell and the resulting vacuum stress $\sigma(r)$ (black curve, in units of $\hbar c/a^4$). As $r\rightarrow a$ the stress $\sigma$ tends to $-\infty$.}
\vspace*{-5mm}
}
\end{center}
\end{figure}

Consider a minor modification of Casimir's model (Fig.~1). Imagine that the spherical shell (though still infinitely conducting and infinitely thin) is no longer hollow, but filled with a medium of gradually varying electric permittivity $\varepsilon$ and magnetic permeability $\mu$. In this way we have extended the shell to a macroscopic body where the Casimir stress gradually builds up. For preserving the spherical symmetry we assume that $\varepsilon$ and $\mu$ depend only on the distance $r$ from the center of the sphere. We expect zero Casimir force in the center, because in a spherically symmetric medium the center does not distinguish any direction for a force vector to point to, and so the force must be zero. The Casimir stress tensor $\sigma$ will be radially symmetric and may change with increasing $r$. Considering how $\sigma$ varies we obtain a physically well-defined Casimir-force density $\nabla \cdot \sigma$. In particular, we assume a toy model for $\varepsilon$ and $\mu$ where $\sigma$ turns out to have an exact solution. Our model is (Fig.~2)
\begin{equation}
\varepsilon = \mu  = \frac{2 n_1}{1+(r/a)^2}
\label{max}
\end{equation}
where $n_1$ is a constant. Equation (\ref{max}) is valid for $r\leq a$, and at $r=a$ we place a perfect spherical mirror that sets the transversal components of the electric field strength to zero. Equation (\ref{max}) describes Maxwell's fish eye \cite{Maxwell} used in perfect imaging \cite{Perfect,Perfect3D,Perfect3Ddiscussion}. We have chosen this model because, in our calculation of the Casimir force, we take advantage of the mathematical fact that Maxwell's fish eye implements the geometry of a simple curved space: it represents the 3-dimensional surface of the 4-dimensional hypersphere in stereographic projection (Fig.~3) \cite{LPDover}. As in other applications of transformation optics \cite{LPDover,CCS} such a geometrical interpretation of electromagnetic media can give us guidance for non-trivial design problems or, as in our case, for calculations that would otherwise be complicated \cite{Einsteinuniverse}. We have found a simple, exact expression for the vacuum-stress tensor:
\begin{equation}
\sigma = - \frac{\hbar c \, \mathds{1}}{\pi^2 a^4 n\, (1-r^2/a^2)^4}
\label{sigma}
\end{equation}
where $\mathds{1}$ is the unity matrix. The stress is isotropic, negative and falls monotonically, so the Casimir-force density $\nabla \cdot \sigma$ is always attractive in our model (Fig.~2). Close to the mirror the stress and the force tend to infinity. Our model thus suggests that the Casimir force at a perfect spherical mirror is indeed infinite. An imperfect mirror, on the other hand, may lead to a finite and possibly attractive vacuum force, which offers new hope for Casimir's fascinating explanation of the finestructure constant and the stability of elementary charged particles \cite{Casimir1}. Of course, in a more realistic theory the particle should not be regarded as being a classical object interacting with the quantum vacuum, as in Casimir's case \cite{Casimir1}, but rather as a self-consistent quantum structure. Speculation aside, we found a non-trivial exact solution for the Casimir force. Analytic solutions for Casimir forces are extremely rare --- the attraction between two plates with infinite $\varepsilon$ \cite{Casimir0}, the repulsion between plates with infinite $\varepsilon$ and $\mu$ \cite{Boyer2} and the attractive force on a homogeneous spherical ball with infinite $\varepsilon$ \cite{MiltonBall} have been solved; we believe we have discovered the first exact solution in a gradually varying medium.

\begin{figure}[t]
\begin{center}
\includegraphics[width=17.0pc]{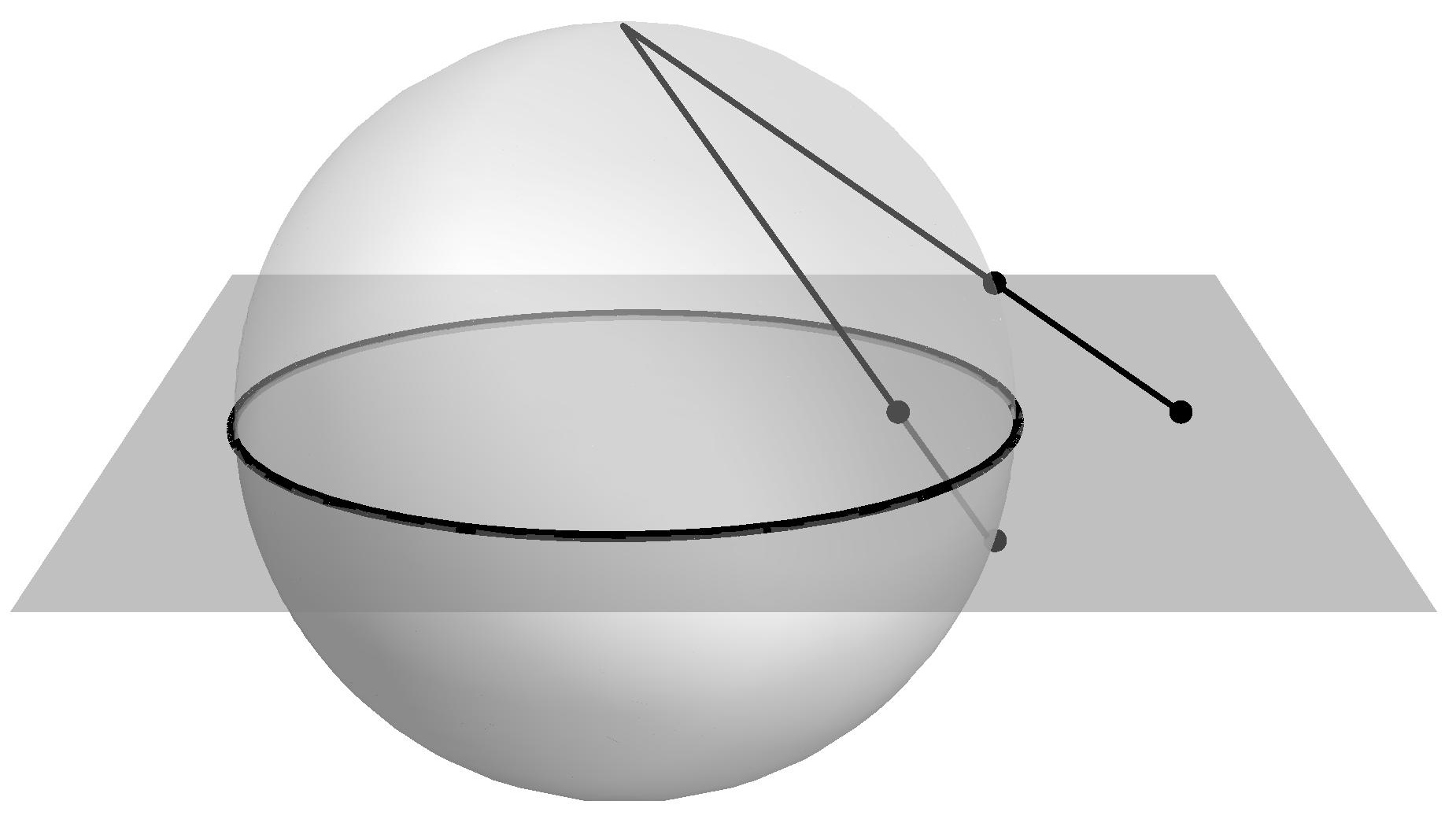}
\caption{
\small{
The medium (plotted in Fig.~2) represents the geometry of the hypersphere (shown here as a sphere) in stereographic projection (lines). The figure illustrates the reflection (points) at the mirror (black circle) on the hypersphere and in physical space.}
\vspace*{-5mm}
}
\end{center}
\end{figure}

We use Lifshitz theory \cite{Leo,Lifshitz} for our calculation, because this is the best physically motivated and tested theory of Casimir forces \cite{Review}. Lifshitz theory relates the vacuum stress to the electromagnetic Green function (as in Schwinger's source theory \cite{Schwinger}). The physical stress $\sigma$ of the quantum vacuum is expressed as
\begin{eqnarray}
\sigma &=& \lim_{\bm{r}_0\rightarrow\bm{r}} \left[\sigma (\bm{r},\bm{r}_0) - \sigma_0 (\bm{r},\bm{r}_0)\right] \,, 
\nonumber \\
\sigma (\bm{r},\bm{r}_0) &=& \tau (\bm{r},\bm{r}_0) - \frac{\mathds{1}}{2} {\rm{Tr}} \, \tau (\bm{r},\bm{r}_0)
\label{corr}
\end{eqnarray}
where the $\tau$ are the correlation functions of the fields in the vacuum state between the points $\bm{r}$ and $\bm{r}_0$ at equal times; the $\tau$ are finite for $\bm{r}\neq\bm{r}_0$. The stress is regularizered by subtracting a bare vacuum stress from $\sigma (\bm{r},\bm{r}_0)$ in the limit $\bm{r}_0\rightarrow\bm{r}$. Our calculation turns out to be independent of the actual regularizer $\sigma_0$ as long as $\sigma_0$ depends only on local properties of the medium. The total correlation function $\tau$ consists of the sum of the electric and the magnetic field correlation functions given by \cite{Leo}
\begin{eqnarray}
\tau_{el} &=& - \frac{\hbar c \, \varepsilon (\bm{r})}{\pi} \int_{0}^{\infty} \kappa^2 G_s (\bm{r}, \bm{r}_0; \mathrm{i} \kappa) \, \mathrm{d} \kappa
\,, \nonumber \\
\tau_{mag} &=& \frac{\hbar c }{\pi \, \mu (\bm{r}_0)} \int_{0}^{\infty} \nabla \times G_s (\bm{r}, \bm{r}_0; \mathrm{i} \kappa) \times \!\stackrel{\longleftarrow}\nabla_0 \, \mathrm{d} \kappa
\label{taudef}
\end{eqnarray}
where the arrow indicates differentiation from the right; $G_s$ denotes the symmetrized electromagnetic Green function $(G + G^{\mathrm T}) /2$ for purely imaginary wavenumbers $\mathrm{i} \kappa$ ({\it i.e.}\ for imaginary frequencies $\mathrm{i} c \kappa$). The Green function $G$ describes the electric field at the spectator point $\bm{r}$ generated by a point source at $\bm{r}_0$ pointing in all possible spatial directions. The Green function is thus a bi-tensor that obeys the wave equation
\begin{equation}
\nabla \times \frac{1}{\mu} \nabla \times G + \varepsilon \kappa^2 G = \delta (\bm{r}-\bm{r}_0) \, \mathds{1} \,.
\label{wave}
\end{equation}
In our case [Eq.~(\ref{max})] the medium is impedance-matched,
\begin{equation}
\varepsilon (\bm{r}) = \mu (\bm{r}) = n (\bm{r}) \,.
\end{equation}
We show that for impedance matching the electric correlation function equals the magnetic one.
For this, we represent $G$ as 
\begin{equation}
G = {\mathrm G} + \frac{\delta (\bm{r}-\bm{r}_0) \, \mathds{1}}{n \kappa^2}
\label{gdef}
\end{equation}
and obtain from the wave equation [Eq.~(\ref{wave})]
\begin{equation}
\nabla \times \frac{1}{n} \nabla \times {\mathrm G} + n \kappa^2 {\mathrm G} = - \frac{\nabla \times \delta (\bm{r}-\bm{r}_0) \, \mathds{1} \times \!\stackrel{\longleftarrow}\nabla_0 }{n (\bm{r}) n (\bm{r}_0) \kappa^2}
\label{wavecurl}
\end{equation}
where we expressed the double curl of the delta-function term [Eq.~(\ref{gdef})] in terms of derivatives with respect to $\bm{r}$ and $\bm{r}_0$. Notice that the magnetic Green function, defined as
\begin{equation}
G_{mag} = - \frac{\nabla \times G \times \!\stackrel{\longleftarrow}\nabla_0 }{n (\bm{r}) n (\bm{r}_0) \kappa^2}  \,,
\label{gmag}
\end{equation}
obeys the same wave equation [Eq.~(\ref{wavecurl})]. Consequently, $G_{mag}$ agrees with $G$ apart from a delta-function term, but such a term does not matter in the correlation functions [Eq.~(\ref{taudef})] where we regard $\bm{r}\neq\bm{r}_0$ before we take the limit $\bm{r}_0\rightarrow\bm{r}$. We conclude that $\tau_{el} = \tau_{mag}$ and obtain
\begin{equation}
\tau = - \frac{2 \, \hbar c \, n}{\pi} \int_{0}^{\infty} \kappa^2 {\mathrm G}_s (\bm{r}, \bm{r}_0; \mathrm{i} \kappa) \, \mathrm{d} \kappa   \,,
\label{tau}
\end{equation}
which sets the scene for calculating the Casimir stress $\sigma$ from the electromagnetic Green function $G$ in impedance-matched media.

The Green function for Maxwell's fish eye [Eq.~(\ref{max})] has been obtained already in connection with perfect imaging \cite{Perfect3D}. We state the results we need here. For keeping our expressions uncluttered we set
\begin{equation}
a = 1 \,,\,\, n_1 = 1
\end{equation}
in our calculation and then obtain the general result [Eq.~(\ref{sigma})] by scaling arguments. Suppose the material [Eq.~(\ref{max})] extends to infinity (without the mirror). In this case \cite{Perfect3Ddiscussion}:
\begin{eqnarray}
{\mathrm G}_0 &=&  - \frac{\nabla \times n (r') \nabla \otimes \nabla_0 \, D (r') \times \!\stackrel{\longleftarrow}\nabla_0 }{n (r) n (r_0) \kappa^2} 
\,,  \label{g0}  \\
r' &=& \frac{|\bm{r}-\bm{r}_0|}{\sqrt{1+ 2\bm{r}\cdot\bm{r}_0 + r^2r_0^2}}
\,,  \label{rmoebius}  \\
D &=& \left( r' + \frac{1}{r'} \right) \frac{\sinh (2 \kappa \, \mathrm{arccot}\,  r')}{8 \pi \sinh (\pi\kappa)}
\,.  \label{d}  
\end{eqnarray}
In the geometrical picture behind Maxwell's fish eye \cite{LPDover} (Fig.~3) the electromagnetic wave propagates on the surface of the hypersphere from source $\bm{r}_0$ to spectator $\bm{r}$ in stereographic coordinates with distance $\mathrm{arctan} \, r'$, and $D$ denotes the Green function of a conformally coupled scalar field \cite{Perfect3Ddiscussion}. The effect of the mirror is described by an adaptation of the method of images \cite{Jackson} on the hypersphere (Fig.~3). There the mirror lies on the equator (a 2-dimensional surface for the 4-dimensional hypersphere). We subtract from ${\mathrm G}_0$ the electromagnetic wave generated by the image source on the hypersphere (Fig.~3). This field is the mirror image of the original field. In stereographic projection \cite{LPDover}, the reflection at the equator corresponds to the transformation $r \rightarrow r^{-1}$. To obtain the reflected wave $\mathrm{G}_0'$, we thus perform the coordinate transformation $\bm{r}=\bm{r}(\bm{r}')$ with $r'=r^{-1}$ and then replace $\bm{r}'$ by $\bm{r}$. Note that we also need to transform the field components of ${\mathrm G}_0$, which is done by the Jacobian \cite{Jacobian}
\begin{equation}
P = \left(\frac{\partial\bm{r}'}{\partial\bm{r}}\right)= \frac{\mathds{1}}{r^2} - \frac{2 \, \bm{r} \otimes \bm{r}}{r^4} 
\label{pdef}
\end{equation}
such that 
\begin{equation}
{\mathrm G}_0' = P \, {\mathrm G}_0 (r^{-1}) \,.
\end{equation}
In this way we obtain for the Green function
\begin{equation}
{\mathrm G} = {\mathrm G}_0 (r) - P \, {\mathrm G}_0 (r^{-1})   \,.
\label{mirrorgreen}
\end{equation}
One verifies that the transversal components of ${\mathrm G}$ vanish at $r=1$, as they should at a perfectly reflecting electric mirror. Note that also for the reflected wave the magnetic Green function equals the electric one (up to an unimportant delta-function term) even if the mirror is not made by an impedance-matched material. To prove this, consider $G_{mag}$ defined by Eq.~(\ref{gmag}) for the transformed Green function ${\mathrm G}_0'$ that describes the reflected wave and rename $\bm{r}$ as $\bm{r}'$ (we recall that ${\mathrm G}_0'$ is the result of a coordinate transformation). Then we make use of two geometrical facts. First, $n(r)^{-1}\nabla\times{\mathrm G}_0$ defines a one-form with respect to the effective geometry with line element $n \, \mathrm{d}l$ \cite{LPDover}. Second, this one-form is invariant under the transformation $r'=r^{-1}$, because the line element $n \, \mathrm{d}l$ is invariant. Therefore we can read $n(r')^{-1}\nabla'\times{\mathrm G}_0'$ as the coordinate-transformed $n(r)^{-1}\nabla\times{\mathrm G}_0$. Hence we can also read  the magnetic Green function of ${\mathrm G}_0'$ (with $\bm{r}$ renamed as $\bm{r}'$) as the coordinate-transformed $G_{mag}$ of ${\mathrm G}_0$. For ${\mathrm G}_0$ the entire medium is impedanced-matched, and in such a case we have already established that the magnetic Green function agrees with the electric one up to a delta-function term. Consequently, the same must be true for the transformed Green function ${\mathrm G}_0'$ that describes the reflection at the mirror.

Now we are ready to calculate the Casimir stress from the electric Green function according to Eqs.~(\ref{corr}) and (\ref{tau}). The Green function ${\mathrm G}_0$ of the infinitely extended fish-eye medium corresponds to the Green function on the entire surface of the hypersphere, which is a uniform space. It can only produce a uniform vacuum stress $\sigma_0$ that does not contribute to the Casimir-force density $\nabla \cdot \sigma$. As we are interested in contributions to $\sigma$ that do generate a force we take the uniform $\sigma_0$ as our regularizer. In this way, we are independent of actual regularization procedures that attempt to explain why the physical vacuum stress is finite. As an additional bonus, we only need to focus on the reflected part of the radiation field, similar to the Lifshitz theory \cite{Leo} for the vacuum stress between conducting plates  \cite{Casimir0} or in other piece-wise uniform planar materials \cite{Lifshitz}. We thus consider only $- P \, {\mathrm G}_0 (r^{-1})$ in the total Green function [Eq.~(\ref{mirrorgreen})]. As the Casimir stress in a spherically symmetric medium must be spherically symmetric, we calculate $- P \, {\mathrm G}_0 (r^{-1})$ only in $x$ direction, {\it i.e.}\ we put $y=z=0$ and $y_0=z_0=0$ in the definitions [Eq.~(\ref{g0}-\ref{pdef})] and evaluate $- P \, {\mathrm G}_0 (r^{-1})$ at $x_0=x^{-1}$. We obtain, after some straightforward algebra,
\begin{equation}
- P \, {\mathrm G}_0 (r^{-1}) = \frac{(1+r^2)^2}{16 \kappa^2 r^4 r'}
\begin{pmatrix}
d_1 & 0 &  0\\
0 & d_2 & 0\\
0 & 0 & d_2
\end{pmatrix}  
\label{pg}
\end{equation}
with the matrix elements
\begin{equation}
d_1 = 2\frac{\mathrm{d}D}{\mathrm{d}r'} ,\quad d_2 = -\frac{\mathrm{d}D}{\mathrm{d}r'} - r' \frac{\mathrm{d}^2 D}{\mathrm{d}{r'}^2}
\label{diff}
\end{equation}
where, in the $x$ direction, $r=x$ and
\begin{equation}
r' = \frac{1}{2} \left( \frac{1}{r} - r \right)  \,.
\label{rprimex}
\end{equation}
Equations (\ref{pg}-\ref{rprimex}) enter the formula for the correlation function [Eq.~(\ref{tau})] in place of ${\mathrm G}_s$ where ${\mathrm G}_s$ is integrated over all positive-imaginary wavenumbers $\mathrm{i} \kappa$. It is wise to perform the $\kappa$ integration before the $r'$ differentiations [Eq.~(\ref{diff})]. We obtain for the scalar Green function [Eq.~(\ref{d})], using integral 2.4.4.1 of Ref.\ \cite{Prudnikov},
\begin{equation}
\int_{0}^{\infty} D \, \mathrm{d} \kappa = \frac{1+{r'}^2}{16 \pi \, {r'}^2} \,,
\end{equation}
which gives 
\begin{equation}
\tau - \tau_0 = \hbar c \, \frac{1+{r}^2}{16 \pi^2 \, (r'r)^4} \, \mathds{1}  \,.
\label{tauresult}
\end{equation}
All three eigenvalues of $\tau - \tau_0$ are identical in the $x$ direction and, by virtue of spherical symmetry, they must be identical in all directions. Equation (\ref{tauresult}) is thus valid everywhere in the medium, and from Eqs.\ (\ref{corr}) and (\ref{rprimex}) we get our result [Eq.~(\ref{sigma})] for $a=1$ and $n_1=1$. Reinstating units for $\sigma$ and $r$ produces Eq.~(\ref{sigma}) for general $a$. For obtaining our result for general $n_1$ we notice that $n_1$ of Eq.~(\ref{max}) appears in the wave equation [Eq.~(\ref{wave})] as $n_1 \kappa$ and a prefactor of $n_1$ of the source term. Consequently, we only need to replace ${\mathrm G} (\mathrm{i} \kappa)$ by $n_1 {\mathrm G} (\mathrm{i} n_1 \kappa)$ in Eq.~(\ref{tau}) where $n$ also carries the prefactor $n_1$, take $n_1 \kappa$ as a new integration variable and obtain Eq.~(\ref{sigma}) in full generality. 

{\it Acknowledgements.---} We thank Simon Horsley, Sahar Sahebdivan and Thomas Philbin for discussions. Our work is supported by EPSRC and the Royal Society.



\begin{thebibliography}{99}
\bibitem{Jacobian}
Note that the Jabobian $P$ given in Ref. \cite{Perfect3D} is incorrect.
\end{thebibliography}
\end{document}